\documentclass[twocolumn,superscriptaddress,aps,preprintnumbers,amsmath,amssymb,prl,nofootinbib]{revtex4-1}

\usepackage[titletoc,toc,title]{appendix}
\usepackage{titletoc}
\usepackage{titlesec}
\usepackage{orcidlink}
\usepackage{textgreek}
\usepackage{amsmath}
\usepackage{amssymb}
\usepackage{amsfonts}
\usepackage{graphicx}
\usepackage{xcolor}
\usepackage{xfrac}
\usepackage{footmisc}
\usepackage{comment}
\usepackage{pifont}
\usepackage{times}
\usepackage{hyperref}
\usepackage{,bm}
\usepackage{enumitem}
\usepackage{centernot}
\usepackage{cancel}
\usepackage{slashed}
\usepackage{relsize}
\usepackage{amsmath,amssymb,mathrsfs}
\usepackage{mdframed}
\usepackage{tcolorbox}
\usepackage{bm}
\usepackage{times}
\usepackage{epsfig}
\usepackage{verbatim}
\usepackage{bm}
\usepackage[utf8]{inputenc}
\usepackage{graphics}
\usepackage{graphicx,epsfig,amssymb,amsmath,color,cancel}
\usepackage{subfigure}
\usepackage[english]{babel}
\usepackage{adjustbox}
\usepackage{soul}
\usepackage[normalem]{ulem}
\usepackage{orcidlink}

\definecolor{zima_blue}{HTML}{1393C1}
\hypersetup{setpagesize=false,bookmarksnumbered=true,bookmarksopen=true,%
colorlinks=true,linkcolor=zima_blue,urlcolor=zima_blue,citecolor=zima_blue,linktocpage=false}

\newtcolorbox{mybox}[3][]{
  colframe = black,         
  colback  = #2!10,
  coltitle = #2!20!black,
  title    = {#3},
  boxsep   = 2pt,           
  top      = 1pt,           
  bottom   = 1pt,           
  #1,
}

\begin{document}

\title{A Universal Bound on the Duration of a Kination Era}

\author{Cem Er\"{o}ncel~\orcidlink{0000-0002-9308-1449}}
\email{cem.eroncel@istinye.edu.tr}
\affiliation{\.{I}stinye University, Faculty of Engineering and Natural Sciences, 34396, \.{I}stanbul, T\"{u}rkiye}
\author{Yann Gouttenoire~\orcidlink{0000-0003-2225-6704}}
\email{yann.gouttenoire@gmail.com}
\affiliation{School of Physics and Astronomy, Tel-Aviv University, Tel-Aviv 69978, Israel}
\affiliation{PRISMA+ Cluster of Excellence $\&$ MITP, Johannes Gutenberg University, 55099 Mainz, Germany}
\author{Ryosuke Sato~\orcidlink{0000-0003-2745-4208}}
\email{rsato@het.phys.sci.osaka-u.ac.jp}
\affiliation{Department of Physics, Osaka University, Toyonaka, Osaka 560-0043, Japan}
\author{G\'{e}raldine Servant~\orcidlink{0000-0002-3063-363X}}
\email{geraldine.servant@desy.de}
\affiliation{Deutsches Elektronen-Synchrotron DESY, Notkestra{\ss}e 85, 22607 Hamburg, Germany}
\affiliation{II. Institute of Theoretical Physics, Universit\"{a}t Hamburg, 22761, Hamburg, Germany}
\author{Peera Simakachorn~\orcidlink{0000-0002-4274-1179}}
\email{peera.simakachorn@ific.uv.es}
\affiliation{IFIC, Universitat de Val\`{e}ncia-CSIC,
C/ Catedr\`{a}tico Jos\'{e} Beltr\`{a}n 2, E-46980, Paterna, Spain}

\begin{abstract}
We show that primordial adiabatic curvature fluctuations generate an instability of the scalar field sourcing a kination era. We demonstrate that the generated higher Fourier modes constitute a radiation-like component dominating over the kination background after about $11$ e-folds of cosmic expansion. Current constraints on the extra number of neutrino flavors $\Delta N_{\rm eff}$ thus imply the observational bound of approximately 10 e-folds, representing the most stringent bound to date on the stiffness of the equation of state of the pre-Big-Bang-Nucleosynthesis universe.

\vspace{0.3cm}
\noindent
\textit{Published version:~\href{https://doi.org/10.1103/k7ty-gwjg}{Phys. Rev. Lett. 135, 101002}}

\end{abstract}

\preprint{DESY-25-010}
\preprint{MITP-25-015}
\preprint{OU-HET-1258}


\maketitle

\section{INTRODUCTION}
The equation of state (EoS) of the universe between the end of inflation and the onset of Big Bang Nucleosynthesis (BBN), defined by the pressure-to-energy density ratio $\omega = p/\rho$, remains unconstrained~\cite{Allahverdi:2020bys}. While the standard model assumes radiation domination with $\omega \simeq 1/3$, a period of \textit{kination}, where the energy density is dominated by the kinetic energy of a fast-rolling scalar field, corresponds to $\omega = 1$~\cite{l1962equation,Spokoiny:1993kt,Joyce:1996cp,Ferreira:1997hj,Joyce:thesis}.
Kination can occur right after inflation in quintessential inflation models~\cite{Peebles:1998qn,Dimopoulos:2001ix}, in string compactifications  \cite{Conlon:2008cj,Conlon:2022pnx,Apers:2024ffe}, or well after reheating during the radiation era, due to a scalar field rotating along a flat direction~\cite{Li:2013nal,Co:2021lkc,Gouttenoire:2021wzu,Gouttenoire:2021jhk,Simakachorn:2022yjy,Harigaya:2023mhl,Harigaya:2023pmw,Chung:2024ctx,Duval:2024jsg}. The latter has been proposed to explain the baryon asymmetry or axion dark matter~\cite{Affleck:1984fy,Co:2019wyp,Co:2019jts,Chang:2019tvx,Co:2020jtv,Eroncel:2022vjg,Eroncel:2022efc,Eroncel:2024rpe}. Kination may also arise from scalar field oscillations in steep potentials $V(\phi) \propto \phi^{p}$ with $p \gg 4$~\cite{Turner:1983he,Poulin:2018dzj,Poulin:2018cxd,Agrawal:2019lmo}, modified gravity~\cite{SanchezLopez:2023ixx}, or small-scale anisotropic stress~\cite{Barrow:1981pa,Turner:1986tc,Niedermann:2019olb,Niedermann:2020dwg}.
With energy density scaling as $\rho \propto a^{-6}$, kination naturally ends when it becomes subdominant to Standard Model (SM) radiation, without requiring decay.  
A well-studied cosmological implication of kination is the enhancement of 
primordial gravitational wave backgrounds (GWBs), generated by inflation~\cite{Sahni:2001qp, Tashiro:2003qp, Sami:2004xk, Artymowski:2017pua, AresteSalo:2017lkv, Figueroa:2018twl,Giovannini:1998bp, Giovannini:2009kg, Giovannini:2022vha,Giovannini:2023wnv,Riazuelo:2000fc,Sahni:2001qp, Seto:2003kc, Tashiro:2003qp,  Sami:2004xk,Boyle:2007zx,Nakayama:2008ip, Nakayama:2008wy, Durrer:2011bi, Kuroyanagi:2011fy, Kuroyanagi:2018csn, Jinno:2012xb, Lasky:2015lej, Li:2016mmc,Artymowski:2017pua,AresteSalo:2017lkv, Saikawa:2018rcs,  Figueroa:2018twl, Figueroa:2019paj,  Caldwell:2018giq, Bernal:2019lpc, DEramo:2019tit,Co:2021lkc,Gouttenoire:2021wzu,Gouttenoire:2021jhk,Servant:2023tua,Li:2013nal, Li:2016mmc, Li:2021htg,SanchezLopez:2023ixx}, cosmic strings~\cite{Cui:2017ufi, Cui:2018rwi, Bettoni:2018pbl,  Auclair:2019wcv, Ramberg:2019dgi, Chang:2019mza,Gouttenoire:2019kij, Gouttenoire:2019rtn,Chang:2021afa,Co:2021lkc,Gouttenoire:2021jhk,Simakachorn:2022yjy,Servant:2023tua}, or first-order phase transitions \cite{Chung:2010cb, Allahverdi:2020bys,Gouttenoire:2021jhk,Servant:2023tua}.
For a large tensor-to-scalar ratio $r$, the enhanced inflationary tensor modes give a large contribution to the number of exotic relativistic degrees of freedom $\Delta N_{\rm eff}$. The Planck bound $\Delta N_{\rm eff}\lesssim 0.34$ ($2\sigma$)~\cite{Planck:2018vyg} converts to an upper limit on the number of kination Hubble expansion e-folding $N_{\rm KD} \lesssim 12.2 +0.5\ln{\left(0.036/r\right)}$, e.g.~\cite{Gouttenoire:2021jhk}.

In this \textit{letter}, we demonstrate that kination not only enhances tensor modes but also amplifies scalar modes, which contribute to $\Delta N_{\rm eff}$ as well. Using $\Delta N_{\rm eff}\lesssim 0.34$ ($2\sigma$), we place the most stringent bound to date on the duration of kination, $N_{\rm KD} \lesssim 10$ e-folds. While the previous bound on $N_{\rm KD}$ from tensor modes strongly depends on the tensor-to-scalar ratio $r$, the new upper bound---derived in this work---is independent of $r$ and solely set by the amplitude of the curvature power spectrum $\mathcal{P}_{\mathcal{R}}(k)$.

\begin{figure}[t!]
\centering
{\makebox{\includegraphics[width=0.475\textwidth, scale=1]{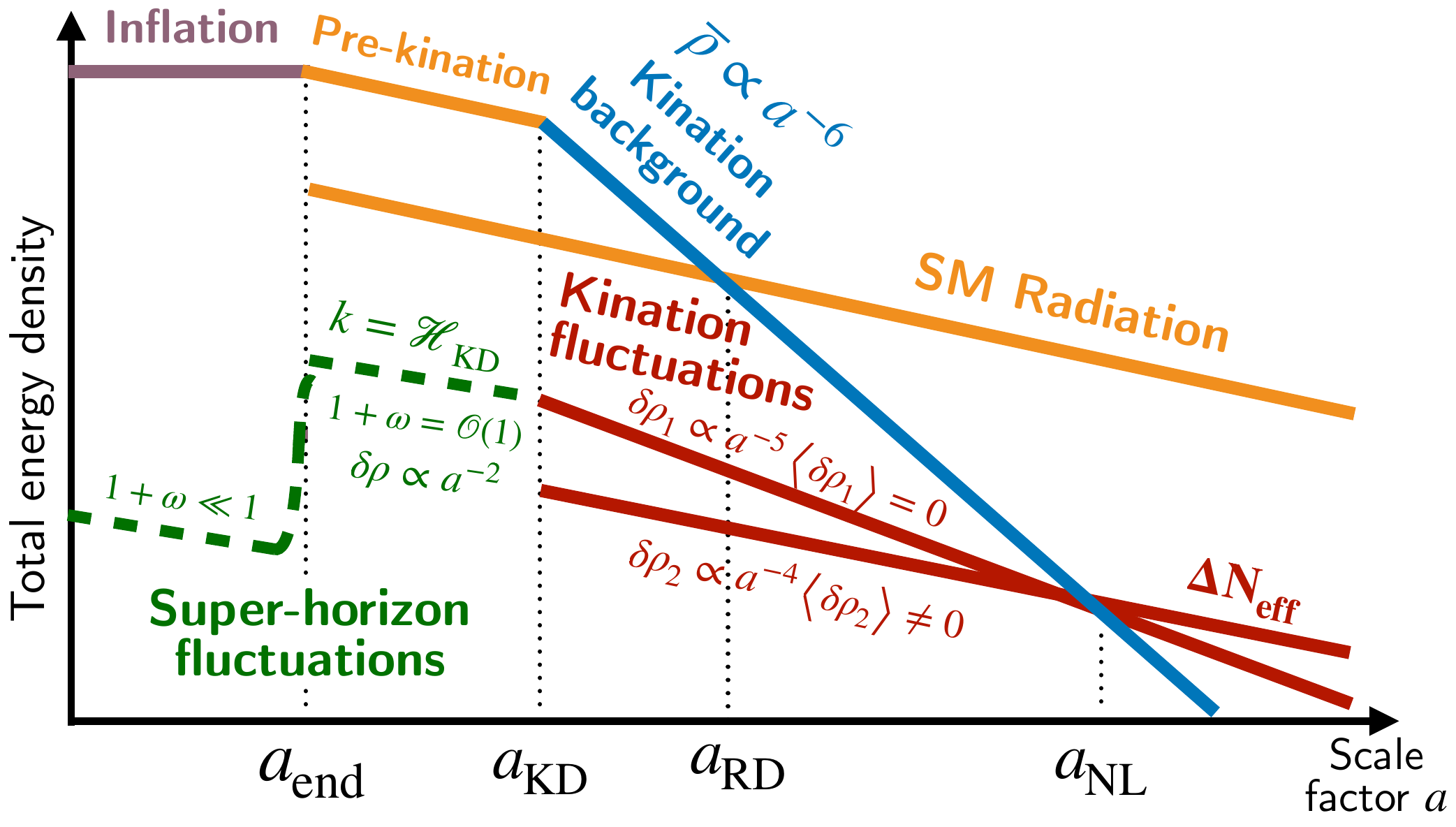}}}
\caption{
During kination, inflationary curvature fluctuations (\textbf{green}) source kination fluctuations (\textbf{red}), which either terminate the kination era if $a_{\rm RD}>a_{\rm NL}$ or contribute to $N_{\rm eff}$ if $a_{\rm RD}<a_{\rm NL}$. The former is excluded by Eq.~\eqref{eq:N_KD_bound_fluctuations}. We consider kination (\textbf{blue}) either immediately after inflation (\textbf{purple}) if $a_{\rm KD}=a_{\rm end}$, as in quintessential inflation scenario \cite{Peebles:1998qn,Dimopoulos:2001ix}, or after unspecified post-inflationary dynamics (\textbf{orange}), which is modeled to be radiation-like for simplicity and a conservative result.
$\delta \rho_1$ and $\delta \rho_2$ are the linear and quadratic corrections to the kination background defined in Eqs.~\eqref{eq:delta_phi_1} and \eqref{eq:delta_phi_2}. The latter implies that kination fluctuations have a non-vanishing volume average and contributes to $\Delta N_{\rm eff}$.
Super-horizon energy fluctuations (green) evolve as $\delta \rho_{\mathbf{k}} \propto (1+\omega)\,\mathcal{R}_{\mathbf{k}}/a^2$, see Eqs.~\eqref{eq:R_vs_Phi} and \eqref{eq:Poisson}. A sudden rise occurs at the end of inflation when $1+\omega = 2\epsilon/3$ initially $\ll 1$ grows to $\mathcal{O}(1)$, with $\epsilon$ being the slow-roll parameter, cf. Eq.~\eqref{eq:epsilon_def}. We picture the super-horizon fluctuations in dashed since they depend on the gauge, here chosen to be the Comoving gauge. We show the mode $k=\mathcal{H}_{\rm KD}$ that re-enters the horizon when kination starts since it is the mode that dominates the contribution to the volume average $\left<\delta \rho_2\right>$, cf. Eqs.~\eqref{eq:rho_fluct_main} and \eqref{eq:Omega_deltaphi}, and therefore to $\Delta N_{\rm eff}$.  The duration of kination is $N_{\rm KD}\equiv \ln(a_{\rm RD}/a_{\rm KD})$. The quantities $a_{\rm end}$, $a_{\rm KD}$, $a_{\rm RD}$, $a_{\rm NL}$ are the scales factors at the end of inflation, start of kination, end of the kination, and start of non-linear regime.
}
\label{fig:Croquis}
\end{figure}

\section{PRIMORDIAL METRIC FLUCTUATIONS}
We consider the cosmological scenario shown in Fig.~\ref{fig:Croquis}. We are interested in the impact of such kination era on the power spectra of primordial curvature $\mathcal{P}_{\mathcal{R}}(k)$ and tensor modes $\mathcal{P}_{t}(k)$.\footnote{The power spectrum of a physical variable $X(\mathbf{r})$ is the Fourier spectrum of the correlation function $P_X(k)=\int d^3\mathbf{r}\, e^{i\mathbf{k}\cdot \mathbf{r}}\left<X(0)X(r) \right>$. In this work, we use the dimensionless power spectrum $ \mathcal{P}_X(k) = \frac{k^3}{2\pi^2}P_X(k)$.}
The amplitude and spectral index of the curvature power spectrum are measured by Planck at the pivot scale $k_{\star}\equiv 0.05~\rm Mpc^{-1}$ to be~\cite{Aghanim:2018eyx},
\begin{equation}
   \mathcal{P}_{\mathcal{R}}(k_{\star})\simeq 2.099(29)\times 10^{-9}, ~ n_s\simeq 0.965(4).
   \label{eq:Delta_R_kstar}
\end{equation}
The non-observation of primordial B-modes in Cosmic Microwave Background (CMB) anisotropies by BICEP/Keck \cite{BICEP:2021xfz} provides the most stringent upper bound on the tensor-to-scalar ratio at pivot scale $k_{\star}$:
\begin{equation}
    r \equiv \mathcal{P}_{t}(k_{\star})/\mathcal{P}_{\mathcal{R}}(k_{\star})\lesssim 0.036,
\end{equation}
leading to a bound on the inflation scale when $k_{\star}$ exit the horizon $H_{\star} \simeq 1.8\times 10^{13}~{\rm GeV}\sqrt{r/0.036}$.
The power spectra at $k \gg k_{\star}$ are unknown and require modeling. In this work, we explore two possibilities. The first considers the curvature and tensor power spectra~\cite{Mukhanov:2005sc},
\begin{equation}
\label{eq:power_spectra_scalar_tensor_slow_roll}
    \mathcal{P}_{\mathcal{R}}(k) = \frac{1}{8\pi^2}\frac{1}{\epsilon_{k}}\left(\frac{H_{k}}{M_{\rm pl}}\right)^2, ~
    \mathcal{P}_{\rm t}(k) = \frac{2}{\pi^2}\left(\frac{H_{k}}{M_{\rm pl}}\right)^2,
\end{equation}
predicted by one of the most strongly supported inflation models~\cite{Aghanim:2018eyx,BICEP:2021xfz,Martin:2024qnn}---the \emph{Starobinsky} inflation~\cite{Starobinsky:1980te,Whitt:1984pd,Ema:2017rqn}, see App.~\ref{app:Starobinsky}.
Here $H_{k}$ and $\epsilon_{k}$ are the Hubble and slow-roll parameters when the comoving scale $k$ exits the horizon.
The second possibility is that the spectra can be well-approximated by power-laws,
\begin{align}
\label{eq:Delta_R_k}
  & \mathcal{P}_{\mathcal{R}}(k) = \mathcal{P}_{\mathcal{R}}(k_{\star})\left({k}/{k_{\star}}\right)^{n_s(k)-1},\\
  \label{eq:Delta_tensor_k}
    & \mathcal{P}_{\rm t}(k)= r\mathcal{P}_{\mathcal{R}}(k_{\star})\left({k}/{k_{\star}}\right)^{n_t(k)},
\end{align}
with the spectral indices $n_s$ and $n_t$ truncated at first-order in $\ln(k)$,
\begin{align}
\label{eq:n_s_k}
    &n_s(k) = n_s(k_{\star})+\alpha_s\ln(k/k_{\star}),\\
 \label{eq:n_t_k}
    &n_t(k) = n_t(k_{\star})+\alpha_t \ln(k/k_{\star}).
\end{align} 
Additionally, the slow-roll assumption leads to the following consistency relations~\cite{Peter:2013avv}:
\begin{align}
    &n_t=-\frac{r}{8}\left(2-n_s-\frac{r^2}{256}\right), ~ \alpha_t=n_t(n_t+1-n_s).
\end{align}
For
$\alpha_s$, we consider the $98\%$ posterior distribution derived from the Bayesian analysis of $\sim$300 models of single-field slow-roll inflation over the latest cosmological data \cite{Martin:2024nlo},
\begin{equation}
\label{eq:alpha_s_posterior}
\alpha_s = -3.65^{+3.20}_{-5.35} \times 10^{-4}.
\end{equation}
We accounted for a factor of $1/2$ due to different definitions of $\alpha_s$ in this work and \cite{Martin:2024nlo}.

\section{$ \Delta N_{\rm eff}$ BOUND FROM TENSOR MODES}%

The already known
upper bound on the duration of kination comes from the enhanced tensor modes, e.g.~\cite{Gouttenoire:2021jhk}.
Primordial tensor modes with dimensionless power spectrum $\mathcal{P}_{\rm t}(k)$ lead to a GWB with energy fraction at late times \cite{Caprini:2018mtu},
\begin{equation}
\Omega_{\rm GW}^{0}(k) = \left(\frac{a_{\mathbf{k}}\mathcal{H}_{\mathbf{k}}}{a_0\mathcal{H}_0} \right)^2  \frac{\mathcal{P}_{\rm t}(k)}{24},
\end{equation}
where $a_{\rm 0}$ and {$a_{\mathbf{k}}$} are the scale factors today and at horizon reentry $k =\mathcal{H}_{\mathbf{k}}$ with  $\mathcal{H}=aH$ being the comoving Hubble factor.
 Modes reentering during the kination era ($H_{k}\propto a_{\mathbf{k}}^{-3}$) yield a slope $k^1$ in the GW spectrum,
\begin{equation}
\label{eq:Omega_GW0_tensor}
    \Omega_{\rm GW}^{0}(k) = \Omega_{\rm rad}^{0}  \mathcal{G}_{k}\frac{\mathcal{P}_{\rm t}(k)}{24}
    \begin{cases}
         k/k_{\rm RD} , &k_{\rm RD}<k<k_{\rm KD},\\
        1, \qquad &k<k_{\rm RD},
    \end{cases}
\end{equation}
where $\Omega_{\rm rad}^0\simeq 9.1\times 10^{-5}$ is today's radiation energy fraction~\cite{ParticleDataGroup:2024cfk}. The quantity $\mathcal{G}_{k}$ accounts for
the changes in relativistic degrees of freedom,
\begin{equation}
  \mathcal{G}_{k}  \equiv\!\left( \frac{g_{\star,k}}{g_{\star,0}} \right)\!\left( \frac{g_{s,0}}{g_{s,k}} \right)^{\!\frac{4}{3}}\!,
\end{equation} 
where $g_\star$ and $g_{s}$ are the numbers of relativistic degrees of freedom contributing to the energy and entropy densities,  respectively. Different subscripts denote the times of today (`$0$'), the mode reentry (`$k$'), the beginning of the kination era (`${\rm KD}$'), and the end of the kination era (`${\rm RD}$').
The contribution of tensor modes to $\Delta N_{\rm eff}$
reads~(cf. App.~\ref{app:BBN_Delta_Neff}):
\begin{equation}
\label{eq:Neff_tensor}
\Delta N_{\rm eff,t} \simeq 0.12 \hspace{-0.1cm}  \int_{k_{\rm RD}}^{k_{\rm KD}}\hspace{-0.1cm} d\log{k}\left(\frac{\mathcal{G}_{k}}{0.39}\right) \left( \frac{k}{k_{\rm RD}}\right) \mathcal{P}_{\rm t}(k),
\end{equation}
where $k_{\rm KD}=\mathcal{H}_{\rm KD}$ and $k_{\rm RD}=\mathcal{H}_{\rm RD}$.
Assuming the power-law $\mathcal{P}_{\rm t}(k)$ in Eq.~\eqref{eq:Delta_tensor_k} with  $N_{\rm KD}\gg1$ and $\mathcal{G}_{k}\simeq 0.39$, Eq.~\eqref{eq:Neff_tensor} becomes, 
\begin{equation}
\label{eq:Neff_tensor_final}
\Delta N_{\rm eff,t} \simeq \frac{0.12}{1+n_t(k_{\rm KD})} \mathcal{P}_{\rm t}(k_{\rm KD})\exp\left(2N_{\rm KD}\right)
\end{equation}
where $N_{\rm KD}$ is the number of e-folds of the kination era,
\begin{equation}
\label{eq:NKD_def}
    N_{\rm KD}\equiv\ln(a_{\rm RD}/a_{\rm KD}).
\end{equation}
Note that $k\propto a^{-2}_k$ during kination era, leading to $N_{\rm KD}=0.5\ln(k_{\rm KD}/k_{\rm RD}
)$.
Using the BAO+CMB constraint $\Delta N_{\rm eff} \lesssim 0.34$ at  $2\sigma$~\cite{Planck:2018vyg}, Eq.~\eqref{eq:Neff_tensor_final} converts into an upper bound on the duration of kination, 
\begin{mybox}{gray}{}\centering{\bf $\Delta N_{\rm eff}$-tensor bound}
\begin{multline}
\label{eq:N_KD_bound_tensor}
    N_{\rm KD} \lesssim 12.2 +0.5\ln{\left(\frac{0.036}{r}\right)} +0.5\ln{\left( \frac{\Delta N_{\rm eff,t}}{0.34}\right)}\\
    - 0.5n_t(k_{\rm KD}) \ln{\left(\frac{k_{\rm KD}}{k_{\star}} \right)}+0.5\ln\left(1+n_t(k_{\rm KD})\right).
\end{multline}
\end{mybox}
A slightly weaker bound can be obtained from BBN, {i.e., $\Delta N_{\rm eff} <0.54$ ($2\sigma$)}  \cite{Pitrou:2018cgg}.
The $N_{\rm KD}$ bound in Eq.~\eqref{eq:N_KD_bound_tensor} depends on the tensor-to-scalar ratio $r$. In a specific inflationary scenario, the value of $r$ could theoretically be much smaller than the BICEP/Keck bound $r \ll 0.036$~\cite{BICEP:2021xfz}, resulting in an upper bound significantly less stringent than Eq.~\eqref{eq:N_KD_bound_tensor}. {Inflaton potentials capable of producing arbitrarily small values of $r$ can be designed~\cite{Brooker:2017vyi,Stein:2022cpk}. Even in top-down constructions such as Kähler moduli inflation, values as low as $r \sim 10^{-10}$ are predicted~\cite{Conlon:2005jm,Martin:2013tda}.}
 We now discuss the effects of kination on scalar perturbations.


\section{GROWTH OF SCALAR MODES DURING KINATION}%

In the absence of anisotropic stress, the metric perturbation in the conformal Newtonian gauge reads~\cite{Mukhanov:2005sc},
\begin{equation}
\label{eq:metric}
    ds^2=a^2\left\{(1+2\Phi)d\eta^2-\left[(1-2\Phi)\delta_{ij}+h_{ij}\right]dx^idx^j\right\},
\end{equation}
where $\Phi$ and $h_{ij}$ are the scalar and tensor fluctuations. 
The gauge-invariant comoving curvature perturbation $\mathcal{R}$ can be expressed in terms of $\Phi$~\cite{Mukhanov:2005sc},
\begin{equation}
\label{eq:R_vs_Phi}
    -\mathcal{R}\equiv \Phi + \frac{2}{3}\frac{(\mathcal{H}^{-1}\Phi'+\Phi)}{1+\omega} \xrightarrow[k\eta \ll 1]{} \frac{5+3\omega}{3+3\omega}\Phi\xrightarrow[\omega=1]{} \frac{4}{3}\Phi.
\end{equation}
In terms of the Fourier mode $\Phi_{\mathbf{k}}=\int d^3\mathbf{x}\,e^{i\mathbf{k}\cdot\mathbf{x}}\Phi({\mathbf{x}})$, the
linear perturbation of Einstein's equations implies~\cite{Mukhanov:2005sc},
\begin{equation}
\label{eq:Phi_ODE}
\Phi_{\mathbf{k}}^{''}(\eta) + \frac{6(1+\omega)}{1+3\omega}\frac{1}{\eta}\Phi'_{\mathbf{k}}(\eta) + \omega k^2 \Phi_{\mathbf{k}}(\eta) = 0,
\end{equation}
The solution of Eq.~\eqref{eq:Phi_ODE} during kination ($\omega = 1$) is:
\begin{equation}
\label{eq:Phik_formula}
    \Phi_{\mathbf{k}}(k\eta) = \frac{2\Phi_{\mathbf{k}}(0)J_1(k\eta)}{k\eta} \xrightarrow[k\eta \gg 1]{}  -\frac{3\mathcal{R}_{\mathbf{k}}(0)}{\sqrt{2\pi}}\frac{\cos(k\eta +\varphi)}{(k\eta)^{3/2}},
\end{equation}
where $\varphi$ is a phase. We deduce the time-averaged scalar power spectrum during kination,
 \begin{equation}
 \label{eq:Delta_phi_0}
     \mathcal{P}_{\Phi}(k,\eta) = \frac{9\mathcal{P}_{\mathcal{R}}(k)}{4\pi(k\eta)^3}.
 \end{equation}
Einstein's equation along 00 leads to Poisson's equation,
 \begin{equation}
 \label{eq:Poisson}
     -k^2 \Phi_{\mathbf{k}} = 4\pi G a^2 \overline{\rho}\,\delta_{\mathbf{k}},
 \end{equation}
 where $\delta_{\mathbf{k}} \equiv ( \rho_{\mathbf{k}} -\overline{\rho})/\overline{\rho}$ is the contrast density in the Comoving gauge~~\cite{Baumann:2022mni}.
Hence, scalar metric fluctuations $\Phi_{\mathbf{k}}(\eta)$ unavoidably generate energy density fluctuations $\delta_{\mathbf{k}}$ {for the dominant background fluid}.
Using 
Eq.~\eqref{eq:Poisson} into Eq.~\eqref{eq:Delta_phi_0} gives rise to the following energy density power spectrum:
\begin{equation}
\label{eq:Delta_delta}
     \mathcal{P}_{\delta}(k,\eta) \simeq \frac{4}{9}\left(\frac{k}{\mathcal{H}}\right)^4 \mathcal{P}_\Phi(k,\eta) =  \frac{8}{\pi}\left(\frac{k}{\mathcal{H}}\right)\mathcal{P}_{\mathcal{R}}(k),
\end{equation}
 where we used Friedmann's equation $3\mathcal{H}^2=8\pi G \overline{\rho} a^2 $, and $\mathcal{H}=2/(1+3\omega)\eta=1/2\eta$ during the kination domination. Using $\eta\propto a^{(1+3\omega)/2}$, we see that fluctuations grow like $\delta\propto a$ during kination. We suppose the kination era to be induced by the kinetic term of massless homogeneous scalar field condensate that gets fragmented into higher modes $\phi =\overline{\phi}+\delta \phi$ such that its energy density
\begin{align}
\label{eq:phi_bkg_energy}
    \overline{\rho}_\phi = \frac{1}{2}\left(\frac{\overline{\phi}'}{a}\right)^2,
\end{align}
receives linear and quadratic corrections $\delta \rho_{\phi} =  \delta \rho_{1}+\delta \rho_{2}$:
\begin{align}
  \label{eq:delta_phi_1}
  &\delta \rho_{1}\simeq {\overline{\phi}^{'}\delta\phi'}/{a^2},\\
  \label{eq:delta_phi_2}
&\delta \rho_{2}= \frac{1}{2}\left(\frac{\delta \phi'}{a}\right)^2+\frac{1}{2}\left(\frac{\nabla \delta \phi}{a}\right)^2\simeq \left(\frac{\delta \phi'}{a}\right)^2,
\end{align}
where we have neglected faster decaying terms containing $\Phi_{\mathbf{k}}$. The gradient term in Eq.~\eqref{eq:delta_phi_2} is equal to the kinetic term since $\delta 
\phi_{\mathbf{k}}\propto \Phi_{\mathbf{k}}(\eta)\propto\cos(k\eta)$. Upon volume averaging the linear term vanishes $\left<\delta\rho_{1}\right>=0$,  while the second term is equal to,
\begin{equation}
\label{eq:rho_fluct_main}
{\left<\delta{\rho_2}\right>}  = \left<\left(\delta\rho_{1}\right)^2/{2\overline{\rho}_\phi}\right>  = \frac{\overline{\rho}_\phi}{2} \int^{k_{\rm KD}}_{k_{\rm RD} }\hspace{-0.2cm}d\log{k} ~\mathcal{P}_{\delta}(k,\eta).
\end{equation}
where we integrate over all fluctuation modes entering during the kination era in the second step.
Plugging $\mathcal{P}_{\delta}(k,\eta)$ in Eq.~\eqref{eq:Delta_delta} into Eq.~\eqref{eq:Omega_deltaphi}, we obtain the fluctuation over background kination energy ratio,
\begin{mybox}{gray}{}
\begin{equation}
\label{eq:Omega_deltaphi}
 \Omega_{\delta_\phi}(k) \equiv \left<\frac{\delta \rho_\phi} {\overline{\rho}_\phi}\right>= \frac{4}{\pi}\int^{k_{\rm KD}}_{k}\hspace{-0.2cm}d\log{\tilde{k}} ~\frac{\tilde{k}}{k}\mathcal{P}_{\mathcal{R}}(\tilde{k})
\end{equation}
\end{mybox}
which we evaluated at the time when $\mathcal{H}(a_k)\equiv k$.
The equation of motion for kination fluctuations is:  
\begin{equation}
\label{eq:EOM_scalar}
\delta\phi_{\mathbf{k}}''+2\mathcal{H}\delta\phi_{\mathbf{k}}' +k^2\delta\phi_{\mathbf{k}} =   4 \Phi_{\mathbf{k}}'\bar{\phi}',
\end{equation}
where the source term is negligible for sub-horizon modes, see App.~\ref{app:fluctuation_radiation}. 
The solution, $\delta \phi \propto a^{-1}$, indicates that kination fluctuations redshift like radiation $\delta \rho_{2}\propto a^{-4}$. {The equation of motion in Eq.~\eqref{eq:EOM_scalar}} includes all orders in $\delta\phi_{\mathbf{k}}$ (only $\Phi_{\mathbf{k}}$ was treated as a perturbation). Hence, the expression for the fluctuation energy density in Eq.~\eqref{eq:EOM_scalar} and the fact that it redshifts like radiation {hold for both} the linear and non-perturbative regimes $\Omega_{\delta_\phi}<1$ and $\Omega_{\delta_\phi}>1$.\footnote{We have checked with $\mathcal{C}$osmo$\mathcal{L}$attice~\cite{Figueroa:2020rrl,Figueroa:2021yhd} that the scalar field energy density redshifts like radiation in the non-linear regime where the gradient energy density becomes comparable to the kinetic energy density. The details will be reported in the companion paper~\cite{companionPAPER}.} 
Note that our derivation of the kination fluctuation relies on the scalar field description, which has been considered in the radiation background in \cite{Eroncel:2022vjg} and kination era in \cite{Apers:2024ffe}. We postpone a full derivation from the perfect fluid formalism for future work.
We now discuss the two novel bounds on the duration of kination which follow from Eq.~\eqref{eq:Omega_deltaphi}.


\begin{figure*}[t!]
\centering
\raisebox{0cm}{\makebox{\includegraphics[width=0.65\textwidth, scale=1]{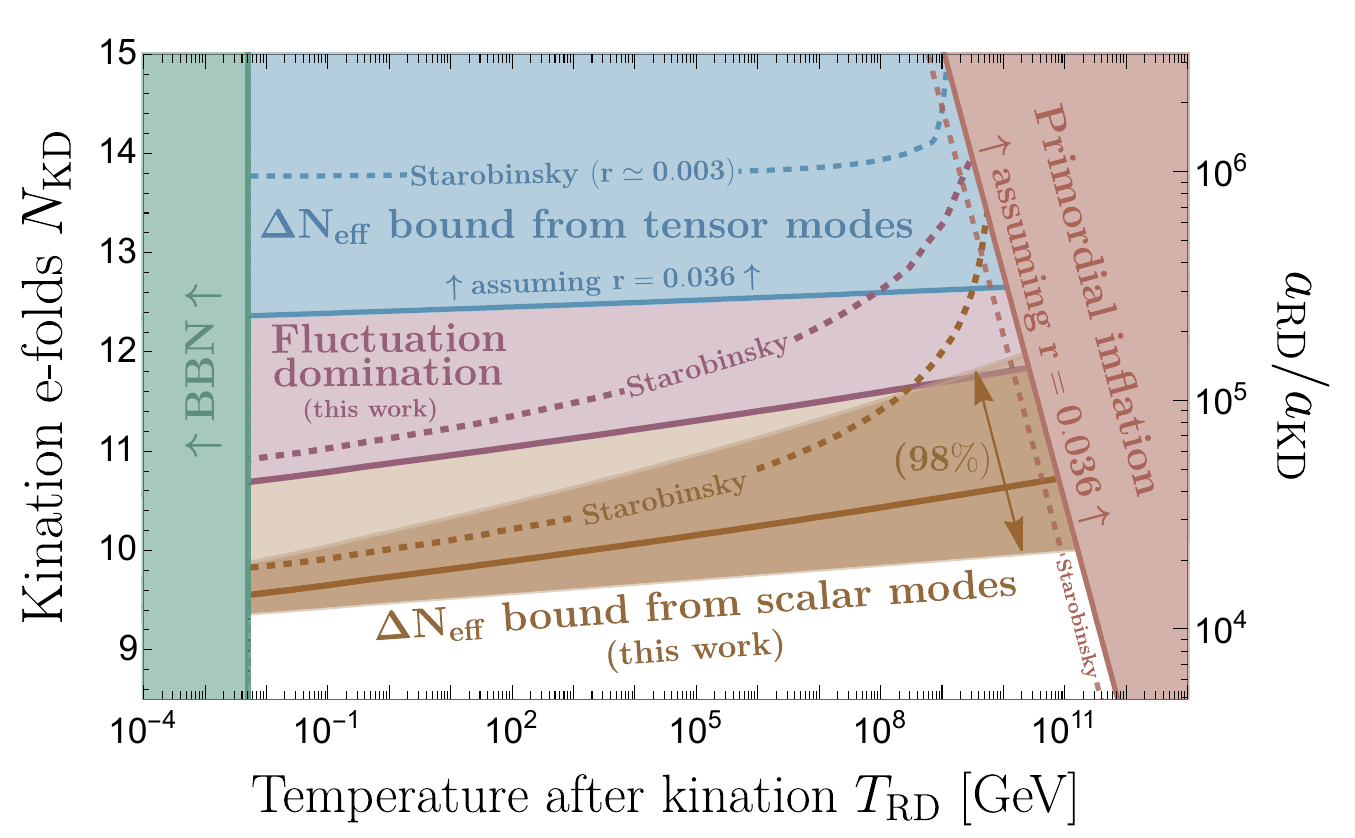}}}
\caption{ Observational bounds on the duration of a kination era due to the linear growth of tensor (Eq.~\eqref{eq:N_KD_bound_tensor} and \textbf{blue} region) and scalar modes (Eq.~\eqref{eq:N_KD_bound_scalar} and \textbf{brown} region) violating the observational contraints $\Delta N_{\rm eff}\lesssim 0.34$ ($2\sigma$)~\cite{Planck:2018vyg}. The \textbf{purple} region is a physical upper limit on the kination duration due to domination by the radiation-like fluctuations, cf. Eq.~\eqref{eq:N_KD_bound_fluctuations}. The power spectra $\mathcal{P}_t(k)$ and $\mathcal{P}_{\mathcal{R}}(k)$ are either given by the power-laws in Eqs.~\eqref{eq:Delta_R_k} and \eqref{eq:Delta_tensor_k} for the \textbf{solid lines} or by the predictions from Starobinsky inflation model in Eq.~\eqref{eq:power_spectra_scalar_tensor_slow_roll} for the \textbf{dashed lines}. The \textbf{brown arrow} indicates the uncertainty in
the spectral-index running $\alpha_s$ in Eq.~\eqref{eq:alpha_s_posterior}, for extrapolating $\mathcal{P}_{\mathcal{R}}(k)$ above the CMB pivot scale $k_{\star}=0.05~\rm Mpc^{-1}$. The solid brown and purple lines assumes $\alpha_s$ to be given by the mean of Eq.~\eqref{eq:alpha_s_posterior}.
In the \textbf{green} region, kination would end after the onset of BBN around $T_{\rm BBN}\simeq 5~\rm MeV$. In the \textbf{red} region, kination would start at a Hubble scale larger than the value $H_{\rm end}$ predicted in Starobinsky inflation (dashed) or its maximal value allowed by the BICEP/Keck bound $r\lesssim 0.036$~\cite{BICEP:2021xfz} (solid). 
} 
\label{fig:New_BBN_bound_powerlaws}
\end{figure*}
\section{FLUCTUATION DOMINATION}%

When $\Omega_{\delta_\phi}(k)>1$ in Eq.~\eqref{eq:Omega_deltaphi}, the energy density of the radiation-like kination fluctuations becomes larger than the energy density of the kination background. This leads to the kination era to terminate and being replaced by a radiation era after a duration $N_{\rm KD}$ which we now determine.
Assuming $\mathcal{P}_{\mathcal{R}}(k)$ to be given by the power-law in Eq.~\eqref{eq:Delta_R_k}, Eq.~\eqref{eq:Omega_deltaphi} becomes for $N_{\rm KD}=0.5\ln(k_{\rm KD}/k_{\rm RD})\gg 1$
\begin{equation}
\label{eq:Omega_deltaphi_2}
\Omega_{\delta_\phi}(k_{\rm RD}) \simeq \frac{4}{\pi n_s(k_{\rm KD})}\mathcal{P}_{\mathcal{R}}(k_{\rm KD})\exp\left(2N_{\rm KD}\right).
\end{equation}
We conclude that the kination era can not physically last longer than
\begin{mybox}{gray}{}\centering{ \bf Fluctuation domination}
\begin{multline}
\label{eq:N_KD_bound_fluctuations}
\!\!\!\!\!\!\!\!\!\!  N_{\rm KD} \lesssim 9.9  +0.5\ln{\left( \frac{2.099 \times 10^{-9}}{\mathcal{P}_{\mathcal{R}}(k_{\star})}\right)}+0.5\ln{n_s(k_{\rm KD})} \\
    + 0.5(1-n_s(k_{\rm KD})) \ln{\left(\frac{k_{\rm KD}}{k_{\star}} \right)}.
\end{multline}
\end{mybox}
This is not an observational constraint but a theoretical bound on the duration of kination, arising from radiation-like fluctuations dominating over the kination background. We now assume that kination terminates before reaching the limit in Eq.~\eqref{eq:N_KD_bound_fluctuations} due to the presence of SM radiation and discuss the contribution of kination fluctations $\Omega_{\delta_\phi}$ to $\Delta N_{\rm eff}$, as illustrated in Fig.~\ref{fig:Croquis}.

\section{$\Delta N_{\rm eff}$ BOUND FROM SCALAR MODES}%

If no entropy injection occurs, the kination fluctuations contribute to the number of effective degree of freedom at the onset of BBN as~(cf. App.~\ref{app:BBN_Delta_Neff}):
\begin{equation}
\label{eq:Neff_scalar}
\Delta N_{\rm eff,s} ~=~ 2.9~\left(\frac{\mathcal{G}_{k}}{0.39}\right)~\Omega_{\delta_\phi}(k_{\rm RD})
\end{equation}
where $\Omega_{\delta_\phi}(k)$ is given by Eq.~\eqref{eq:Omega_deltaphi}.
The constraint $\Delta N_{\rm eff} \lesssim 0.34$ at  $2\sigma$~\cite{Planck:2018vyg} can be recast as:
\begin{mybox}{gray}{}\centering{ \bf $\Delta N_{\rm eff}$-scalar bound}
\begin{multline}
\label{eq:N_KD_bound_scalar}
\!\!\!\!\!\!\!\!\!\!  N_{\rm KD} \lesssim 8.8 +0.5\ln{\left( \frac{\Delta N_{\rm eff,s}}{0.34}\right)} +0.5\ln{\left( \frac{2.099 \times 10^{-9}}{\mathcal{P}_{\mathcal{R}}(k_{\star})}\right)} \\
    + 0.5(1-n_s(k_{\rm KD})) \ln{\left(\frac{k_{\rm KD}}{k_{\star}} \right)}+0.5\ln{n_s(k_{\rm KD})}.
\end{multline}
\end{mybox}
The bound in Eq.~\eqref{eq:N_KD_bound_scalar} is more restrictive than both Eq.~\eqref{eq:N_KD_bound_tensor} and Eq.~\eqref{eq:N_KD_bound_fluctuations}. 
We compare the three bounds in  Fig.~\ref{fig:New_BBN_bound_powerlaws}.
 In contrast to the already-known bound in Eq.~\eqref{eq:N_KD_bound_tensor}, the two novel bounds in Eqs.~\eqref{eq:N_KD_bound_scalar} and \eqref{eq:N_KD_bound_fluctuations} are independent of the -- unmeasured -- value of the tensor-to-scalar ratio $r$. 

\section{CONCLUSION}

We have studied how cosmological perturbations evolve during a kination era. At linear order, fluctuations grow linearly with the scale factor, $\delta \propto a$ like matter fluctuations in agreement with previous findings --- see e.g.~\cite{Redmond:2018xty,Eroncel:2022vjg}. Accounting for non-linear corrections, we demonstrated that fluctuations of the scalar driving kination have a non-vanishing volume average $\langle \delta \rangle \neq 0$ cf. Eq.~\eqref{eq:rho_fluct_main}, unlike matter fluctuations due to mass conservation~\cite{Peebles:1980yev,Bernardeau:2001qr}, and redshift like radiation. Consequently, a kination era induced by a scalar field must end after approximately 11 e-folds when these fluctuations become non-perturbative and induce a radiation-like equation of state (EoS) --- see the purple curves in Fig.~\eqref{fig:New_BBN_bound_powerlaws}.\footnote{Prior to our work, the instability of kination had been demonstrated in the context of oscillating scalar condensates~\cite{Lozanov:2017hjm} where the fluctuations arise from the fragmentation of
the scalar field rather than being sourced by curvature perturbations.} 

These radiation-like fluctuations also contribute to the effective number of neutrino species $\Delta N_{\rm eff}$. Planck $2\sigma$ limit, $\Delta N_{\rm eff}\lesssim 0.34$~\cite{Planck:2018vyg}, currently imposes the strongest constraint on kination, restricting its duration to $N_{\rm KD} \lesssim 10$ e-folds --- see {Eq.~\eqref{eq:N_KD_bound_scalar} and} the brown curves in Fig.~\eqref{fig:New_BBN_bound_powerlaws}.
Unlike the already-known bound from tensor modes in Eq.~\eqref{eq:N_KD_bound_tensor}, the new limits from fluctuation domination and scalar modes in Eqs.~\eqref{eq:N_KD_bound_fluctuations} and \eqref{eq:N_KD_bound_scalar} derived in this work are independent of the unknown value of the tensor-to-scalar ratio $r$. Future measurements, such as CMB-S4 with a projected sensitivity of $\Delta N_{\rm eff} \lesssim 0.05$~\cite{CMB-S4:2016ple,Dvorkin:2022jyg}, could further tighten the bounds on $N_{\rm KD}$ by about one e-fold.

 {The conservative bounds presented in this work rely on the assumption that the scalar field acquires the primordial curvature perturbations from the inflaton, which requires that the scalar field either inherits its energy density from the inflaton or is the inflaton itself, as demonstrated in Fig.~\ref{fig:Croquis}.}\footnote{{In addition to those originating from adiabatic curvature fluctuations, kination fluctuations could also arise from isocurvature perturbations, which, however, depend on the scalar field dynamics during inflation and are left for future study.}}
 
We considered for simplicity scenarios where kination is preceded by either a radiation era or inflation \cite{Peebles:1998qn}. 
Scalar field theories leading to a radiation EoS preceding kination are discussed in \cite{SanchezLopez:2023ixx}. On the other hand, kination in rotating axion models~\cite{Co:2021lkc,Gouttenoire:2021wzu,Gouttenoire:2021jhk,Simakachorn:2022yjy,Harigaya:2023mhl,Harigaya:2023mhl,Harigaya:2023pmw,Chung:2024ctx,Duval:2024jsg}, which is disconnected from inflation and preceded by a matter era,
 would amplify scalar perturbations more strongly, leading to a more stringent $\Delta N_{\rm eff}$ bound.  Our bound should therefore be seen as conservative. Moreover, if fluctuations become massive due to the scalar field condensate acquiring a mass, the constraints derived in this work would be further tightened, or the scalar fluctuations could account for dark matter~\cite{Eroncel:2025qlk,Bodas:2025eca}. 
 We leave for future works the study of kination fluctuations starting from the stress-energy tensor of a perfect fluid instead of the Lagrangian of a scalar field. This would, allow to possibly extrapolate our results to any eras with stiff EoS $\omega>1/3$.

{\bf Acknowledgements.}---%
We are grateful to Konstantinos Dimopoulos and Pedro Schwaller for stimulating discussions.
YG acknowledges support by the Cluster of Excellence ``PRISMA+'' funded by the German Research Foundation (DFG) within the German Excellence Strategy (Project No. 390831469), and by a fellowship awarded by the Azrieli Foundation. 
PS is supported by Generalitat Valenciana Grants:  PROMETEO/2021/083 and CIPROM/2022/69. This work is also supported by the Deutsche Forschungsgemeinschaft under Germany’s Excellence Strategy---EXC 2121 ``Quantum Universe"---390833306.
The work of RS is supported in part by JSPS KAKENHI Grant Numbers~23K03415, 24H02236, and 24H02244.

\appendix
\onecolumngrid

\fontsize{11}{13}\selectfont




\newpage
\vspace{1cm}

\begin{center}
\textbf{\it\Large Supplemental Material
}
\end{center}
\noindent

\renewcommand{\tocname}{\large  Contents
\vspace{-0.2 cm}}%

   \titleformat{\section}
  {\normalfont\fontsize{12}{14}\bfseries  \centering }{\thesection:}{1em}{}
  \titleformat{\subsection}
  {\normalfont\fontsize{12}{14}\bfseries \centering}{\thesubsection.}{1em}{}
  \titleformat{\subsubsection}
  {\normalfont\fontsize{12}{14}\bfseries \centering}{\thesubsubsection)}{1em}{}
  
     \titleformat{\paragraph}
  {\normalfont\fontsize{12}{14}\bfseries  }{\thesection:}{1em}{}

{
  \hypersetup{linkcolor=black}

}

\section{Assuming a specific inflation model}
\label{app:Starobinsky}
CMB measurements from Planck and BICEP/Keck~\cite{Aghanim:2018eyx,BICEP:2021xfz,Martin:2024qnn} 
indicate that plateau-like concave inflation potentials are favored over the convex-type potentials. One such potential is the $R+R^2$ model proposed by Starobinsky~\cite{Starobinsky:1980te}:
\begin{equation}
    S=\int d^4x\sqrt{-g}\left[\frac{M_{\rm pl}^2}{2}R+\frac{\xi_s}{16} R^2\right],
\end{equation}
where $R$ is the scalar curvature and $\xi_s$ is some coupling constant. This theory is conformally equivalent to Einstein gravity with a scalar field $\phi$ with potential~\cite{Whitt:1984pd,Ema:2017rqn}:
\begin{equation}
    V(\phi) = M_{\rm pl}^4 \xi_s^{-1}\left[1-e^{-\frac{\phi}{M_{\rm pl}}\sqrt{\frac{2}{3}}}\right]^2.
\end{equation}
The slow-roll parameter $\epsilon$ reads:
\begin{align}
\label{eq:epsilon_def}
    &\epsilon\equiv \frac{\dot{\phi}^2}{2M_{\rm pl}^2H^2}\simeq \frac{M_{\rm pl}^2}{2}\left(\frac{V'}{V}\right)^2=\frac{4}{3}\left[1-e^{-\frac{\phi}{M_{\rm pl}}\sqrt{\frac{2}{3}}}\right]^{-2}.
\end{align}
Inflation ends when the slow-roll condition ($\epsilon < 1$) breaks down,
\begin{align}
\label{eq:epsilon_end}
    \epsilon(\phi_{\rm end}) = 1 ~ \Rightarrow ~ \phi_{\rm end} =\sqrt{\frac{3}{2}}M_{\rm pl}\ln\left(1+\frac{2}{\sqrt{3}}\right).
\end{align}
The number of remaining inflation e-folds $N_k\equiv \ln\left( a_{\rm end}/a_{k}\right)$ when the cosmological scale $k=a_k H_{k}$ leaves the horizon can be written as:
\begin{equation}
     N_k =-\ln\left(\frac{k}{k_{\rm KD}}\right)
     +\ln\left(\frac{\mathcal{H}_{\rm end}}{\mathcal{H}_{\rm KD}}\right)-\ln\left(\frac{H_{\rm end}}{H_{k}}\right).
\end{equation}
Assuming the kination era follows a radiation era as shown in Fig.~\eqref{fig:Croquis}, we have $\mathcal{H}_{\rm end}/\mathcal{H}_{\rm KD}=a_{\rm KD}/a_{\rm end}$.
The number of inflation e-folds before the inflation ends is,
\begin{equation}
    N_k \equiv \ln\left( \frac{a_{\rm end}}{a_{\mathbf{k}}}\right)= \int_{\phi_{\rm inf}}^{\phi_{k}} \frac{H}{\dot{\phi}} d\phi=\int_{\phi_{\rm end}}^{\phi_{k}} \frac{d\phi}{M_{\rm pl}}\frac{1}{\sqrt{2\epsilon(\phi)}},
\end{equation}
which upon inverting gives $\phi_{k}(N_k)$. 
We deduce the curvature and tensor power spectra~\cite{Mukhanov:2005sc}:
\begin{equation}
\label{eq:power_spectra_scalar_tensor_slow_roll_app}
    \mathcal{P}_{\mathcal{R}}(k) = \frac{1}{8\pi^2}\frac{1}{\epsilon_{k}}\left(\frac{H_{k}}{M_{\rm pl}}\right)^2,\quad
    \mathcal{P}_{\rm t}(k) = \frac{2}{\pi^2}\left(\frac{H_{k}}{M_{\rm pl}}\right)^2,
\end{equation}
where $H_{k}^2=V(\phi_{k})/(3M_{\rm pl}^2)$ and $\epsilon_{k}$ is obtained from plugging $\phi_{k}(N_k)$ into Eq.~\eqref{eq:epsilon_def}. For every kination duration $N_{\rm KD}$ and temperature  $T_{\rm RD}$, we use Eq.~\eqref{eq:Delta_R_kstar} to fix the constant $\xi_s$, and calculate the contribution of tensor and scalar modes to $\Delta N_{\rm eff}$ via Eqs.~\eqref{eq:Neff_tensor} and \eqref{eq:Neff_scalar}. The $2\sigma$ Planck bound $\Delta N_{\rm eff}\lesssim 0.34$~\cite{Planck:2018vyg} leads to the constraints shown in the blue dotted and brown dotted lines in Fig.~\ref{fig:New_BBN_bound_powerlaws}.

\section{$\Delta N_{\rm eff}$ bound}
\label{app:BBN_Delta_Neff}

GWs or higher modes of the scalar field act as \emph{dark radiation} with energy density today $\rho_{\rm DR}(T_0)$, which contributes to the number of neutrino flavors, defined by
\begin{equation}
\label{eq:Delta_Neff}
 \Delta N_{\rm eff}(T_0) = \frac{8}{7}\left( \frac{11}{4} \right)^{4/3}\!\!\left( \frac{\rho_{\rm DR}(T_0)}{\rho_\gamma(T_0)} \right),
\end{equation}
where $\rho_{\gamma}$ represents the photon number density and $T_0$ refers to any temperature below the neutrino decoupling threshold $T_0\lesssim \rm MeV$.
The value of $\Delta N_{\rm eff}$ is limited by both BBN~\cite{Pitrou:2018cgg} and CMB~\cite{Planck:2018vyg,Dvorkin:2022jyg} measurements to be $\Delta N_{\rm eff}  \lesssim 0.34$. Substituting  $\Omega_{\rm DR}(T_{k})=\rho_{\rm DR}(T_{k})/\rho_{\rm rad}(T_{k})$ into Eq.~\eqref{eq:Delta_Neff}, we arrive at,
\begin{align}
    \Delta N_{\rm eff}    ~ = ~ \frac{8}{7}\left( \frac{11}{4} \right)^{\!4/3}\!\!\left( \frac{g_{\star}(T_{k})}{g_\gamma} \right)\!\left( \frac{g_{\star,s}(T_0)}{g_{\star,s}(T_{k})} \right)^{\!4/3}\!\!\!\!\Omega_{\rm DR}(T_{k})~\simeq ~2.89 ~\left(\frac{\mathcal{G}_{k}}{0.39}\right) ~ \Omega_{\rm DR}(T_{k}),
    \label{eq:Delta_Neff_bound}
\end{align}
with $ \mathcal{G}_{k}  \equiv\!\left( g_{\star,k}/g_{\star,0} \right)\!\left( g_{s,0}/g_{s,k} \right)^{\!4/3}\!$.
The quantity $g_\gamma=2$ corresponds to the number of photon degrees of freedom, $g_\star$ and $g_s$ are the number of relativistic degrees of freedom in the energy and entropy densities, with the values today $g_{\star,0} = 2+(7/8)\cdot 2\cdot N_{\rm eff}\cdot (4/11)^{4/3} \simeq 3.38$, and $g_{s,0} = 2+(7/8)\cdot 2\cdot N_{\rm eff}\cdot (4/11) \simeq 3.94$, assuming  $N_{\rm eff}\simeq 3.043$ \cite{Cielo:2023bqp}. For $g_{\star,k}=g_{s,k}=106.75$, we get $ \mathcal{G}_{k}\simeq 0.39$.
{We take the evolution of $g_\star$ and $g_s$ from \cite{Saikawa:2018rcs}.}
Eq.~\eqref{eq:Neff_tensor} in the main text is obtained by replacing $\Omega_{\rm DR}(T_{k})$ in Eq.~\eqref{eq:Delta_Neff_bound} by $ \Omega_{\rm GW}^{0}(k)/ \Omega_{\rm rad}^{0}$ in Eq.~\eqref{eq:Omega_GW0_tensor}. We divide by $ \Omega_{\rm rad}^{0}$ because we are interested in $\Delta N_{\rm eff}$ before matter-radiation equality. 

In the analysis of the main text, we treat $\Delta N_{\rm eff}$ from tensor and scalar separately. In fact, $\Delta N_{\rm eff}= \Delta N_{\rm eff,t}
+\Delta N_{\rm eff,s}$ where $\Delta N_{\rm eff,t}$ and $\Delta N_{\rm eff,s}$ are given in Eqs.~\eqref{eq:Neff_tensor_final} and~\eqref{eq:Neff_scalar}. We can write: 
\begin{align}
    \frac{\Delta N_{\rm eff,t}}{\Delta N_{\rm eff,s}} \simeq 0.03\,r(k_{\rm KD})\left(\frac{n_s(k_{\rm KD})}{1+n_t(k_{\rm KD})}\right)\simeq 0.5\,\epsilon(k_{\rm KD})\left(\frac{n_s(k_{\rm KD})}{1+n_t(k_{\rm KD})}\right),
\end{align}
where we used $r(k)\equiv \mathcal{P}_t(k)/\mathcal{P}_\mathcal{R}(k)\simeq 16\epsilon(k)$, following from Eq.~\eqref{eq:power_spectra_scalar_tensor_slow_roll_app}. We conclude that the $\Delta N_{\rm eff}$-scalar bound is stronger than the $\Delta N_{\rm eff}$-tensor bound for modes $k$ exiting the horizon during the slow roll phase $\epsilon(k)\ll 1$. However, tensor and scalar bound become comparable for modes exiting the horizon near the end of inflation when the slow-roll parameter becomes large $\epsilon(k) \sim 1$ as indicated by Eq.~\eqref{eq:epsilon_end}.

\section{Kination fluctuations behave like radiation}
\label{app:fluctuation_radiation}

In this appendix, we use the scalar field equations of motion to show that kination fluctuations redshift like radiation both in the perturbative and non-perturbative regime.
At first order in $\Phi$, the Lagrangian of the scalar field can be written:
\begin{equation}
L
= \sqrt{-g} \left( \frac{1}{2} g^{\mu\nu}(\partial_\mu \phi)(\partial_\nu \phi) \right)
\simeq a^4 \left( \frac{1}{2a^2}(1 - 4\Phi)\phi^{'2} - \frac{1}{2a^2}(\nabla \phi)^2 \right),
\end{equation}
where we have used the metric given in Eq.~\eqref{eq:metric}.
Using Euler-Lagrange equation for $\phi$, we derive the equation of motion:
\begin{equation}
\phi'' + 2{\cal H}\phi' - \nabla^2\phi = 4\Phi(\phi'' + 2{\cal H}\phi') + 4\Phi' \phi'.
\end{equation}
The Fourier transform of the Newtonian potential can be determined from Eq.~\eqref{eq:Phi_ODE} as (for $\omega>0$)
\begin{equation}
    \Phi_{\mathbf{k}} = \Phi_{\mathbf{k}}(\eta=0) \times \Gamma(\alpha+1) \left( \frac{2}{\sqrt{\omega}k\eta} \right)^{\alpha} J_\alpha(\sqrt{\omega} k\eta) ,\qquad \textrm{where}\quad \alpha \equiv \frac{5+3\omega}{2(1+3\omega)}.\label{eq:PHI after horizon entry}
\end{equation}
Hence, $\Phi$ is a decreasing function and always remains perturbative. Thus, $4\Phi(\phi'' + 2{\cal H}\phi')$ in RHS is negligible compared to $\phi'' + 2{\cal H}\phi'$ in LHS and we can simplify the EOM as
\begin{equation}
\phi'' + 2{\cal H}\phi' - \nabla^2\phi = 4\Phi' \phi'.
\end{equation}
Let us expand $\phi$ as $\phi=\overline{\phi}+\delta \phi$. Then, the Fourier transform of the above EOM is
\begin{equation}
\label{eq:EOM_scalar}
\overline{\phi}'' + 2\mathcal{H}\overline{\phi}' = {\cal S}_0,\qquad
\delta\phi_{\mathbf{k}}''+2\mathcal{H}\delta\phi_{\mathbf{k}}' +k^2\delta\phi_{\mathbf{k}} = \mathcal{S}_{\mathbf{k}},
\end{equation}
where
\begin{equation}
\mathcal{S}_{\mathbf{0}} = 4\int \frac{d^3\mathbf{p}}{(2\pi)^3}\Phi_{\mathbf{p}}'\delta \phi_{-\mathbf{p}}', \qquad
\mathcal{S}_{\mathbf{k}}=4 \Phi_{\mathbf{k}}'\overline{\phi}'+ 4\int \frac{d^3\mathbf{p}}{(2\pi)^3}\Phi_{\mathbf{p}}'\delta \phi_{\mathbf{k}-\mathbf{p}}',
\end{equation}
After transient behavior at horizon entry due to the source term, $\delta\phi_{\mathbf{k}}$ quickly becomes the solution of the homogeneous equation for which $\Phi_{\mathbf{k}}$ is set to zero:
\begin{equation}
    \delta \phi_{\mathbf{k}}(\eta) = \eta^{-\alpha+1} \left[A_{\mathbf{k}}J_{\alpha-1}(k\eta)+B_{\mathbf{k}} Y_{\alpha-1}(k\eta)\right],
\end{equation}
where $J(x)$ and $Y(x)$ are the Bessel functions of the first and second kinds. We deduce that in the deep sub-horizon limit, the scalar fields behave as
\begin{equation}
\lim_{k\eta \to +\infty}\delta \phi_{\mathbf{k}}(\eta) = \sqrt{\frac{2}{k\pi}} \eta^{-\alpha+1/2} \left[A_{\mathbf{k}}\cos\left(k\eta -\frac{\pi}{2}(2\alpha-1)\right)+B_{\mathbf{k}}\sin\left(k\eta -\frac{\pi}{2}(2\alpha-1)\right)\right].
\end{equation}
Using that $a\propto\eta^{\alpha-1/2}$, we deduce 
\begin{equation}
\label{eq:delta_phi_app_radiation}
    \delta \phi_{\mathbf{k}}(\eta) ~\propto \frac{1}{k^{1/2}}\frac{\cos\left(k\eta\right)}{a},~ {\rm for} ~ k\eta\gg 1.
\end{equation}
We deduce that $\delta\phi'_{\mathbf{k}}\xrightarrow[k\eta \to +\infty]{} -k^{-1}\delta\phi_{\mathbf{k}}$.
Hence, the energy density of the fluctuation becomes
\begin{equation}
\label{eq:delta_rho_app_radiation}
    \delta \rho_{2}= \frac{1}{2}\left(\frac{\delta \phi'}{a}\right)^2+\frac{1}{2}\left(\frac{\nabla \delta \phi}{a}\right)^2~\simeq~ \left(\frac{\delta \phi'}{a}\right)^2 ~\propto~k \frac{\cos^2(k\eta)}{a^4}.
\end{equation}
We conclude that the time-averaged energy density of the kination fluctuation redshifts  like radiation $\delta \rho_{2}\propto a^{-4}$ independently of the background EoS $\omega$. Note that we have recovered the spectral enhancement  $\delta \rho_{2}\propto k^1$ in Eq.~\eqref{eq:Omega_deltaphi}. It is important to note that Eq.~\eqref{eq:EOM_scalar} is valid at all orders in $\delta \phi_{\mathbf{k}}$. We conclude that the expression in Eqs.~\eqref{eq:delta_phi_app_radiation} and \eqref{eq:delta_rho_app_radiation} are valid both in the linear $\delta\phi_{\mathbf{k}}/\overline{\phi}<1$ and non-linear regime $\delta\phi_{\mathbf{k}}/\overline{\phi}>1$.  Using $\mathcal{C}$osmo$\mathcal{L}$attice~\cite{Figueroa:2020rrl,Figueroa:2021yhd}, we have run lattice simulations to check that fluctuations interpreted as the gradient and kinetic terms in Eq.~\eqref{eq:delta_phi_2} redshift like radiation whether they are dominating or not~\cite{companionPAPER}.

We now check a posteriori that it was a good approximation to neglect the source term $\mathcal{S}_{\mathbf{k}}$ in Eq.~\eqref{eq:EOM_scalar} in the sub-horizon limit.
The Newtonian potential during kination and radiation era reads $\Phi_{\mathbf{k}}\propto a^{-3}\cos(k\eta)$ and $a^{-2}\cos(k\eta/\sqrt{3})$ respectively, cf. Eq.~\eqref{eq:Phi_ODE}. Using $\overline{\phi}'\propto a^{-2}$ and $\delta \phi_\mathbf{k}'\propto a^{-1}$, we obtain that the source term $\mathcal{S}_{\mathbf{k}}$ in Eq.~\eqref{eq:EOM_scalar} redshifts as 
\begin{equation}
    \mathcal{S}_{\mathbf{k}}\propto \begin{cases}
        a^{-3},\qquad \textrm{for }\omega=1,\\
        a^{-2},\qquad \textrm{for }\omega=\frac{1}{3},\\
    \end{cases}
\end{equation}
which is faster than the left-hand side of Eq.~\eqref{eq:EOM_scalar} which redshifts as $k^2\delta\phi_{\mathbf{k}}\propto a^{-1}$. We conclude that we can safely neglect the effect of $\Phi_{\mathbf{k}}$ on the evolution of $\delta \phi_{\mathbf{k}}$ for sub-horizon modes.

\bibliography{biblio}

\end{document}